\documentclass[preprint,prb,aps]{revtex4}

\usepackage{graphicx}
\usepackage{color}
\usepackage{textcomp}
\usepackage{mathrsfs, amsmath, amssymb}

\begin{document}

\title{Transition-Metal Pentatelluride ZrTe$_5$ and HfTe$_5$:
  a Paradigm for Large-gap Quantum Spin Hall Insulators }

\author{Hongming Weng}


\author{Xi Dai}

\email{daix@aphy.iphy.ac.cn}

\author{Zhong Fang}

\email{zfang@aphy.iphy.ac.cn}

\affiliation{Beijing National Laboratory for Condensed Matter
  Physics, and Institute of Physics, Chinese Academy of Sciences,
  Beijing 100190, China;}

\date{\today}

\begin{abstract}

  Quantum spin Hall (QSH) insulators, a new class of quantum matters,
  can support topologically protected helical edge modes inside
  bulk insulating gap, which can lead to dissipationless
  transport. A major obstacle to reach wide application of QSH is the
  lack of suitable QSH compounds, which should be easily fabricated and
  has large size of bulk gap. Here we predict that single layer 
  ZrTe$_5$ and HfTe$_5$ are the most promising candidates to reach 
  the large gap QSH insulators with bulk direct
  (indirect) band gap as large as 0.4 eV (0.1 eV), and robust against
  external strains. The 3D crystals of these two materials are good 
  layered compounds with very weak inter-layer bonding and are  
  located near the phase
  boundary between weak and strong topological insulators, which pave
  a new way to future experimental studies on both QSH effect and
  topological phase transitions.

\end{abstract}

\maketitle

\section{Introduction}
Topological insulators (TI)~\cite{TI-1,TI-2} can support gapless
boundary states inside bulk insulating gap, which are topologically
protected and robust against perturbations. Particularly in
two-dimensional (2D) TIs, namely the QSH
insulators~\cite{Kane,Bernevig}, the low energy (back) scattering of
the edge states is prohibited by the time reversal symmetry, leading
to the dissipationless transport edge channels and the QSH
effect~\cite{QSH-theory}. However, the known experimental
verifications of QSH effect in HgTe/CdTe~\cite{QSH-HgTe} and
InAs/GaSb~\cite{QSH-InAs} quantum well structures require extreme
conditions, such as the precisely controlled molecular-beam-epitaxy
(MBE) and the ultra low temperature (due to the small bulk band gap of
the order of meV), which greatly obstruct the further experimental
studies and possible applications.

To be a "good" QSH insulator, the material must meet the following
important criteria: (1) It must be a good layered materials to get the
2D system easily; (2) It must have large 2D bulk band gap to realize
the QSH effect at high temperature.  The first proposal for the QSH
insulator in graphene~\cite{Kane} is practically useless due to its
extremely small gap (10$^{-3}$meV) induced by spin-orbit coupling
(SOC)~\cite{Yao}, although graphene is so far the best 2D material
which can be made easily even by scotch
tape~\cite{Graphene-scotch}. Recently Bi${_2}$TeI has been proposed 
to be another candidate for QSH insulator\cite{bi2TeI}, but its energy gap is also small. 
The Bismuth (111)-bilayer is potentially
a large-gap (about 0.2 eV) QSH insulator~\cite{Bi}, however, it is
inter-layer strongly bonded and its fabrication is hard (not
experimentally achieved yet). Several other proposals, such as the
ultra thin tin films~\cite{Tin} and the Cd$_3$As$_2$ quantum
well~\cite{Cd3As2}, have the same problem. Since they are not layered
materials, the well controlled MBE technique is required to obtain
the ultra thin film samples.

Here we report that simple binary ZrTe$_5$ and HfTe$_5$, previously
known as layered thermoelectric compounds, are inter-layer weakly
bonded comparable to that of graphite. Their single layer sheets,
which can be made in principle with no need for MBE, are QSH
insulators with large bulk gap, and robust against the lattice
distortions. Therefore, ZrTe$_5$ and HfTe$_5$ are the most
promising 2D TI, which satisfy both the above conditions and pave a new way to
more experimental studies on QSH effect. Moreover, our calculations
show that their three dimensional (3D) crystals formed by the stacking of 
layers are located in the vicinity of a transition between strong and weak TI, 
which further makes it a perfect platform to study the topological
quantum phase transitions.

\section{Computational Methods}

The {\it ab initio} calculations have been done by using the all
electron full potential linearized augmented plane wave method
implemented in the WIEN2k package\cite{wien2k} within the general
gradient approximation (GGA). The choice of parameters, such as
sampling of Brillouin Zone and cut-off of augmented plane waves, are
carefully checked to ensure the convergence. The spin-orbit coupling
(SOC) is included self-consistently within the second variational
method. The inter-layer binding energy, i.e., the unit area
total-energy difference between the single layer sheet and the 3D
bulk, is calculated by using the local density approximation (LDA) of
Perdew and Wang type exchange correlation~\cite{pw92}. It is known
that LDA can give the Van der Waals type inter-layer coupling energy
reasonably well~\cite{interlayer}. The maximally localized Wannier
functions (MLWF) for Te $p$ orbitals have been constructed by using
home made code~\cite{mlwf} based on OpenMX~\cite{openmx}, which can
reproduce the band structures calculated from the first-principles
quite well. The surface and edge states have been calculated for slabs
by using a tight-binding model constructed from the MLWFs generated
above.

\section{Results}
\subsection{Layered crystal structure} 
Both ZrTe$_5$ and HfTe$_5$ take the same structure and have very
similar properties. They have been shown to possess interesting
electrical transport properties, such as the giant resistivity
anomaly~\cite{anomaly,anomaly-by-H} and the large
thermopower~\cite{thermo, M-doping}. They crystallize in the
orthorhombic layered structure~\cite{structure} with space group
$Cmcm$ ($D_{2h}^{17}$) as shown in Fig. 1 (hereafter we take ZrTe$_5$
as an example). Trigonal prismatic chains of ZrTe$_3$ (marked by red
dashed line) run along the $a$ axis, and these prismatic chains are
linked via parallel zigzag chains of Te atoms along the $c$ axis to
form 2D sheet of ZrTe$_5$ in the $a$-$c$ plane. The sheets of ZrTe$_5$
stack along the $b$ axis, forming layered structure. The prism of
ZrTe$_3$ is formed by a dimer of Te$^d_{1,2}$ atoms (with superscript
$d$ indicating the dimer and subscript $1,2$ numbering the atoms) and
one apical Te$^a$ atom (with $a$ meaning the apical), while the zigzag
chain is formed by two Te$^z_{1,2}$ atoms (with $z$ indicating the
zigzag chain).  The primitive unit cell contains two formula units with 
two prisms and two zigzag chains. The corresponding subscripts numbering
the atoms should be doubled.

\begin{figure}[tbp]
\includegraphics[clip,scale=0.5]{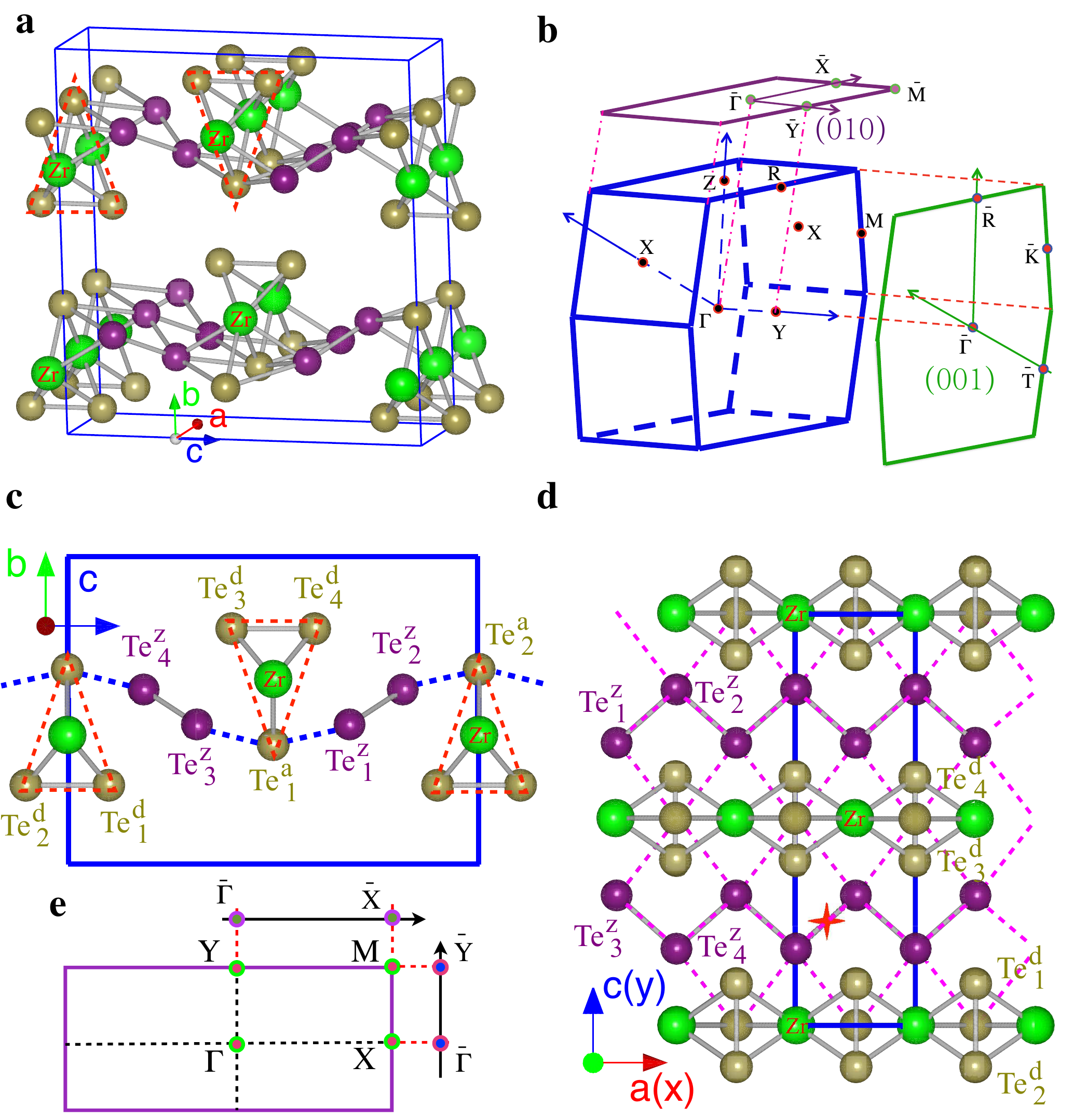}
\caption{
  (a) The crystal structure and (b) the bulk
  Brillouin Zone (BZ) and the projected surface BZ of 3D ZrTe$_5$
  (HfTe$_5$). (c), (d) and (e) is the side view, top view, and BZ of
  single layer structure, respectively. In (d), the inversion center
  is indicated by the red star symbol, and the waved grid of Te square
  lattice sheet is shown as the pink dotted lines. 
 } 
 \end{figure}

ZrTe$_5$ shows strong quasi-2D anisotropy~\cite{anisotropy}. The prism
chains and zigzag chains are connected through the apical Te atoms,
and the Te-Te bond length between two chains is just about 0.4 \AA\
longer than that in the zigzag chain. If the Zr and Te$^d$ dimer atoms
are neglected, the remaining apical Te$^a$ and zigzag Te$^z$ atoms can
be viewed as a waved grid of Te square lattice sheet (see Fig.1(d)),
leading to a stable quasi-2D structure. Each ZrTe$_5$ layer is
nominally charge neutral, and the inter-layer distance (along $b$
axis) is quite large (about 7.25 \AA), suggesting the weak inter-layer
coupling of presumably Van der Waals type. We have calculated the
inter-layer binding energies for different compounds by using the
first-principles total energy method (see method).  The results shown
in Fig. 2 suggest that the inter-layer binding energy of ZrTe$_5$ or
HfTe$_5$ is as weak as that of graphite, and is much smaller than that
of Bi$_2$Se$_3$ and Bi (111)-bilayer. Given the easy procedure of
making graphene from graphite simply by scotch
tape~\cite{Graphene-scotch}, the comparably weak inter-layer binding
energy suggests that the single layer of ZrTe$_5$ (or HfTe$_5$) may be
formed in a similar simple and efficient method.

\begin{figure}[tbp]
\includegraphics[clip,scale=0.6]{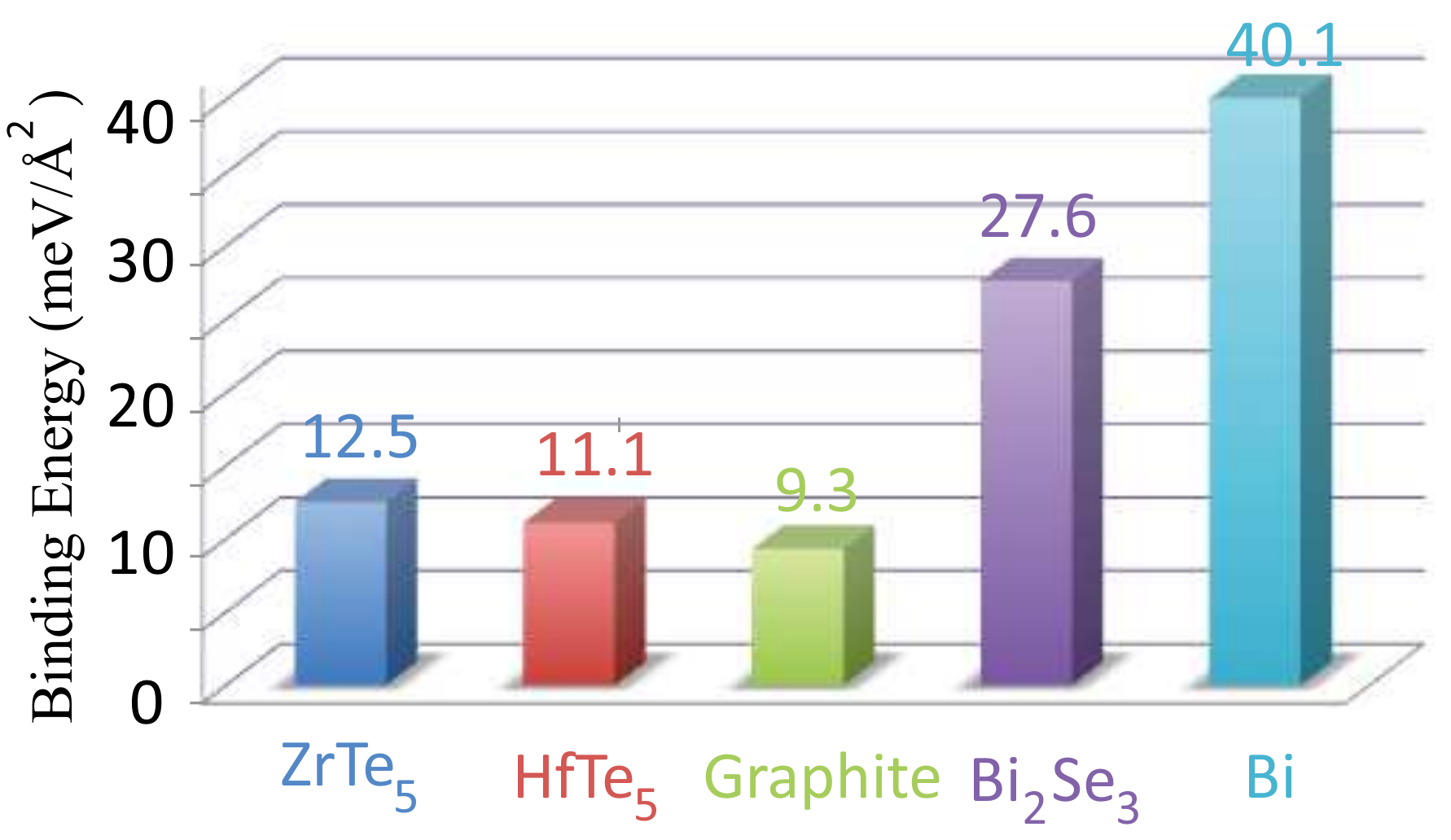}
\caption{
 The calculated inter-layer binding energies
    for several typical layered compounds.
 }
 \end{figure}

\subsection{ 2D Quantum Spin Hall Insulator} 
We now focus on the single layer of ZrTe$_5$ in the $a$-$c$ plane.
Its structure is fully optimized by theoretical calculation. It is
found that the relaxed lattice constants ($a$=4.036 \AA\ and
$c$=13.843 \AA) and the internal atomic coordinates are all very close
to their 3D bulk experimental values (less than one percent
difference)~\cite{structure}, again suggesting the stability of the
single layer sheet. The calculated density of states shown in
Fig. 3(a) suggest that the Zr-$4d$ states are mostly unoccupied above
the Fermi level, leading to the nearly ionic Zr$^{4+}$ state (by
transferring $\sim$4 electrons onto the neighboring Te atoms). The
states near the Fermi level are dominantly from the covalently bonded
Te-$5p$ states.

\begin{figure}[tbp]
\includegraphics[clip,scale=0.5]{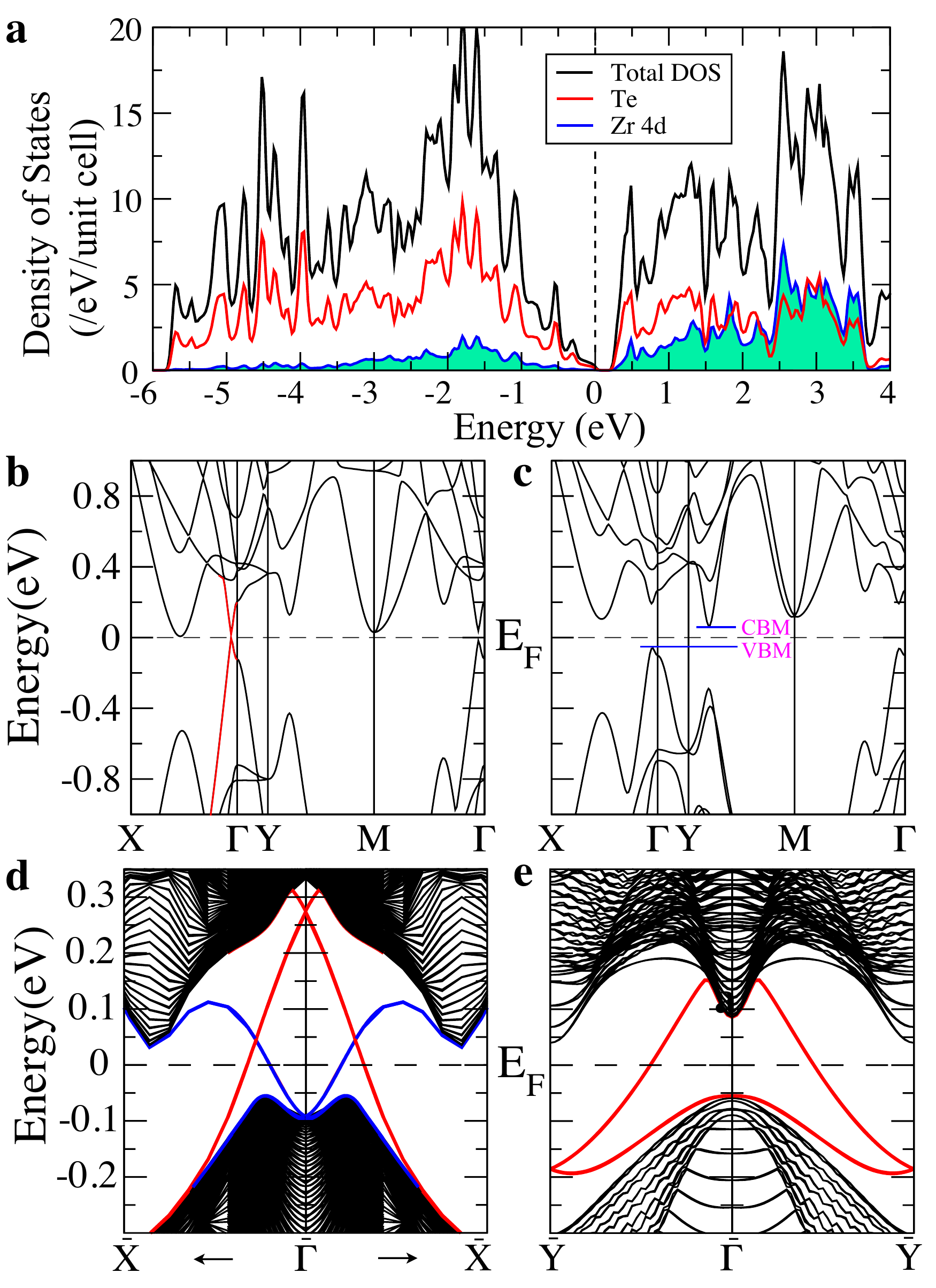}
\caption{
 The calculated electronic structures of single
  layer ZrTe$_5$. (a) The density of states; (b) and (c) the band
  structures without and with SOC, respectively. The valence band
  maximum (VBM) and conduction band minimum (CBM) defining the
  in-direct band gap is indicated in (c); (d) and (e) are the calculated
  edge states for the $x$- and $y$-edges, respectively (see the main
  text for details).
  }
\end{figure}

For the convenience of discussions, hereafter we define our coordinate
system with $x$, $y$ being along the $a$ and $c$-axis, respectively,
and $z$ being out of the 2D plane. We choose the origin of the coordinate system
to be located on the Zr site.
The calculated band structures for a single
layer ZrTe$_5$ are shown in Fig. 3 (b) and (c). In the case without
SOC, the system is a semimetal with a band-crossing along the
$\Gamma$-X direction, implying the existence of band inversion around
the $\Gamma$. The band crossing is unavoidable because the two bands
belong to different representations distinguished by the $m_{xz}$
symmetry along the $\Gamma$-X axis. The inclusion of SOC will,
however, mix them and open up a gap, resulting in an insulator with
large direct (indirect) band gap of 0.4 eV (0.1 eV).  We have check
the $Z_2$ topological invariant~\cite{Kane} of the resulting
insulating state by evaluating the parity eigen values of occupied
states at four time-reversal-invariant-momenta (TRIM) points of the
Brillouin Zone (BZ)\cite{parity}, and conclude that it is a non-trivial QSH
insulator with $Z_2$=1. Since the band gap is mostly determined by SOC
strength, we wouldn't expect much error bar of GGA type calculations
(as it usually does for conventional semiconductors). Indeed, we have
checked the electronic structure by using the hybrid functional
HSE06,~\cite{HSE} as well as the modified Becke-Johnson (mBJ)
potential~\cite{mBJ}. Both of them give the similar band gap with the
band topology unchanged.

The 2D non-trivial insulating state in single layer ZrTe$_5$ should
support topologically protected conducting edge states. Two idea edges 
along the $x$- and $y$-direction are studied by tight-binding
slab models (see method). The termination is
between the prism chain and the zigzag chain for the former, and is
between the two Te$^z$ atoms of the zigzag chain for the later. The
results are shown in Fig. 3(d) and (e). For the case of cutting along
the $x$ direction, the left and right edges of our slab are not
symmetric, which leads to two separated Dirac cones at $\bar{\Gamma}$:
one for the edge with prism chain termination (red), and other one for
the zigzag chain termination (blue). While for the case of cutting
along the $y$ direction, both sides of the slab are symmetric and two
Dirac cones (located at opposite sides) are energetically
degenerate. There is another difference between $x$- and $y$-edge.
The Dirac cones are at $\bar{\Gamma}$ in the former, while at
$\bar{Y}$ in the later.

\subsection{Band inversion mechanism} 
The band inversion in single layer ZrTe5
is not due to SOC, instead it is mainly 
due to the non-symmorphic features of the  space group. Therefore the physics of 
band inversion here can be understood without SOC, and the only effect of SOC
is to open a energy gap afterwards.  
We note that the space group of the 
system is $P_{mmn}$ ($D_{2h}^{13}$), which is a non-symmorphic one with 
the inversion center located not at the origin but at (1/4,1/4).
 As an important consequence of such space group symmetry, the eigenstates at all zone
boundary TRIM points (namely, X, Y, and M) are all four-fold degenerate
(including spin degrees of freedom) with two of them having even
parity and the other two odd parity. This is because that at all the three zone boundary points
the inversion operation anti-commutes with two important mirror operators 
$m_{xz}$ and $m_{yz}$. (see detail in Appendix). 
The equal number of parities guarantees that
any band inversion at those zone boundary TRIM points will not change
the topology of the system and the $Z_2$ index of the material is
fully determined by the energy order of the bands at the $\Gamma$
point, where only the Kramer degeneracy holds.  We further notice that
the mirror symmetry $m_{yz}$ will transfer all the atoms into
themselves. Therefore, the atomic $p$ orbitals can be classified into two
classes, namely, the $p_x$ orbital and the $p_{y/z}$ orbital, because they will not
mix in the eigenstates of $\Gamma$ point (in the absence of SOC).

Our calculations suggest that the band inversion at $\Gamma$ happens
between the zigzag chain Te$^z$-$p_x$ and the prism chain dimer
Te$^d$-$p_{y}$ states, as shown schematically in Fig. 4. At the atomic
limit, they are all four-fold degenerated since there are four
equivalent Te atoms for each class. The strong intra-chain covalence
bonding will split them into double degenerate bonding and
anti-bonding states, respectively, and then the weak inter-chain
coupling will further split them into singly degenerate states. It is
important to note the symmetry difference between the two chains. The
two dimer Te$^d_{1,2}$ atoms in the same prism chain are related by
the mirror $m_{xz}$ symmetry, while the two Te$^z_{1,2}$ atoms in the same
zigzag chain are related by the inversion symmetry. The bonding and
antibonding states caused by the intra-chain coupling can therefore be
distinguished by the eigenvalues $m_{xz}$=$\pm 1$ for the former, and
the parity $p$=$\pm 1$ for the later. Due to the strong intra-chain
coupling, the band inversion is introduced between the bonding Te$^d$
states with $m_{xz}$=1 and the antibonding Te$^z$ states with $p$=-1,
as illustrated in Fig. 4. Within the two bonding Te$^d$ states of
$m_{xz}=1$, only one state has odd parity. As a result, its occupation 
changes the total parity of the occupied states at $\Gamma$, which
leads to the QSH state.

\begin{figure}[tbp]
 \includegraphics[clip,scale=0.5]{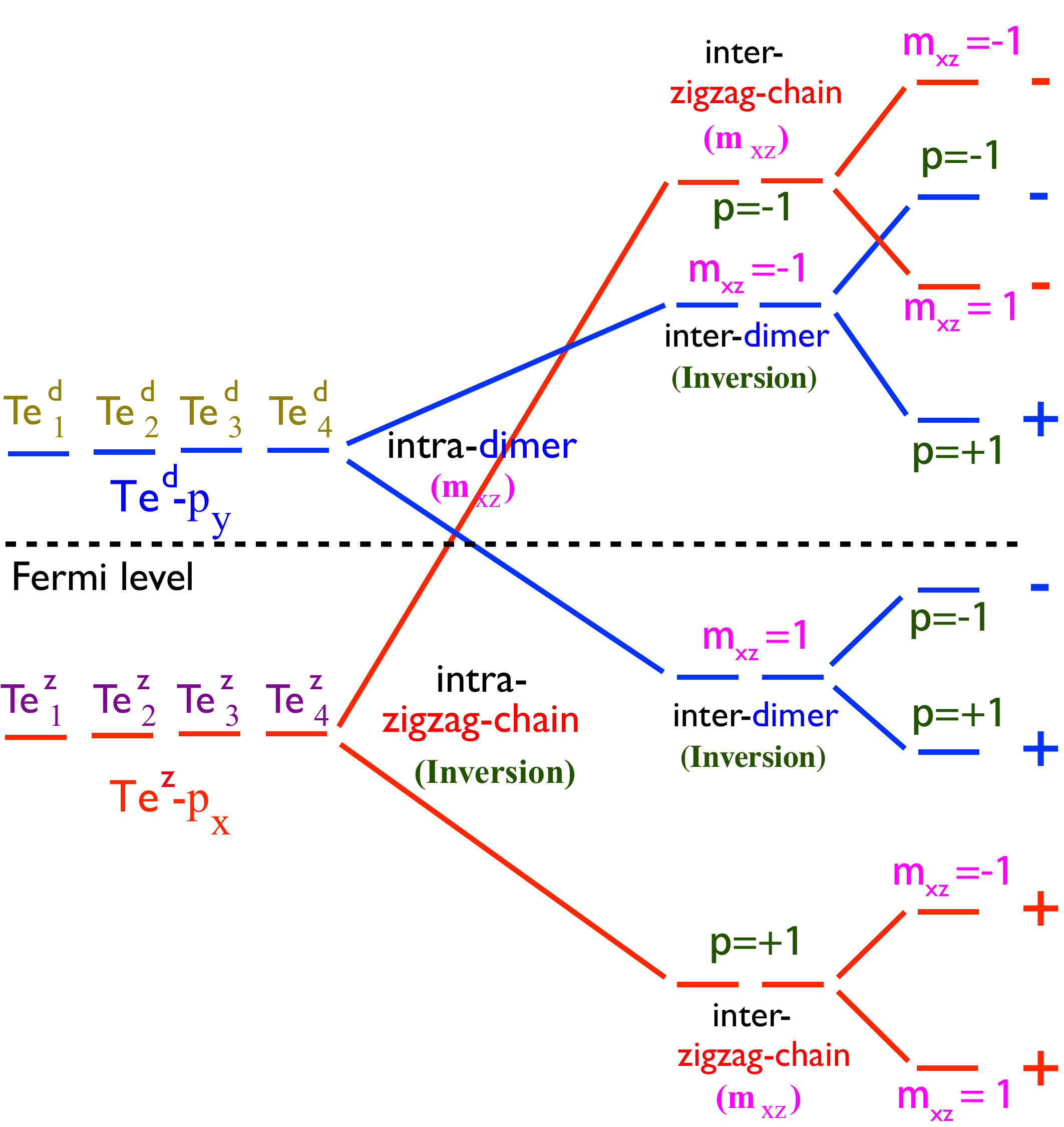}
 \caption{
 The schematic illustration of the band
    inversion mechanism (see main text for details).
}
\end{figure}

\subsection{Strain effects} 
We emphasize that, unlike the situation in typical TI such as the
Bi$_2$Se$_3$ family compounds~\cite{Bi2Se3}, the band inversion in
ZrTe$_5$ single layer is not due to SOC, instead it is due to the
special non-symmorphic space group features as discussed above.  As
long as the inter-chain coupling is not strong enough to reverse the
band ordering again (between the ($m_{xz}$=-1, $p$=+1) state and the
($m_{xz}$=1, $p$=-1) state, as shown in Fig. 4), the QSH state should
be stable. We have checked the stability of this QSH state as function
of strains. Without lost of generality, here we assume that the
lattice parameters $a$ and $c$ are equally scaled by possible external
potentials, and for each fixed volume we fully optimize the internal
coordinates. The results shown in Fig. 5 suggest that the QSH state
survives even if the volume is expanded by more than 20\% or
compressed by 10\%, indicating its robust stability against the
strains. This makes it highly adaptable in various application
environments.  In the case of compressing, the too much enhanced
inter-chain coupling will finally kill the QSH state, as discussed
above.

\begin{figure}[tbp]
  \includegraphics[clip,scale=0.5]{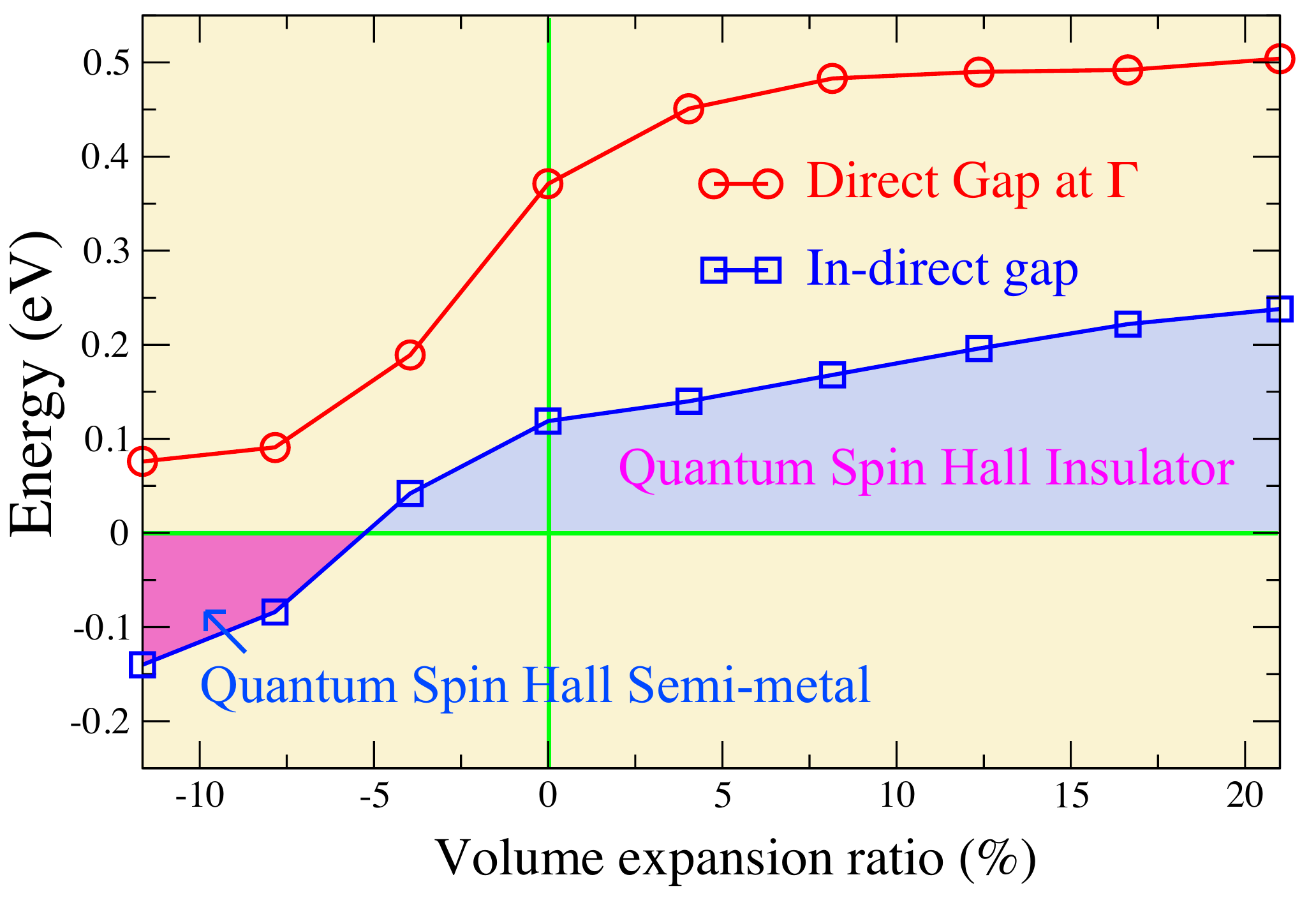}
  \caption{
   The calculated band gaps of single layer
    ZrTe$_5$ as function of volume change. Both the direct gap at
    $\Gamma$ and the in-direct band gap are shown. The non-trivial
    $Z_2$ topology survives as long as the band gap at $\Gamma$
    remains positive.
    }
\end{figure}

\begin{figure}[tbp]
 \includegraphics[clip,scale=0.5]{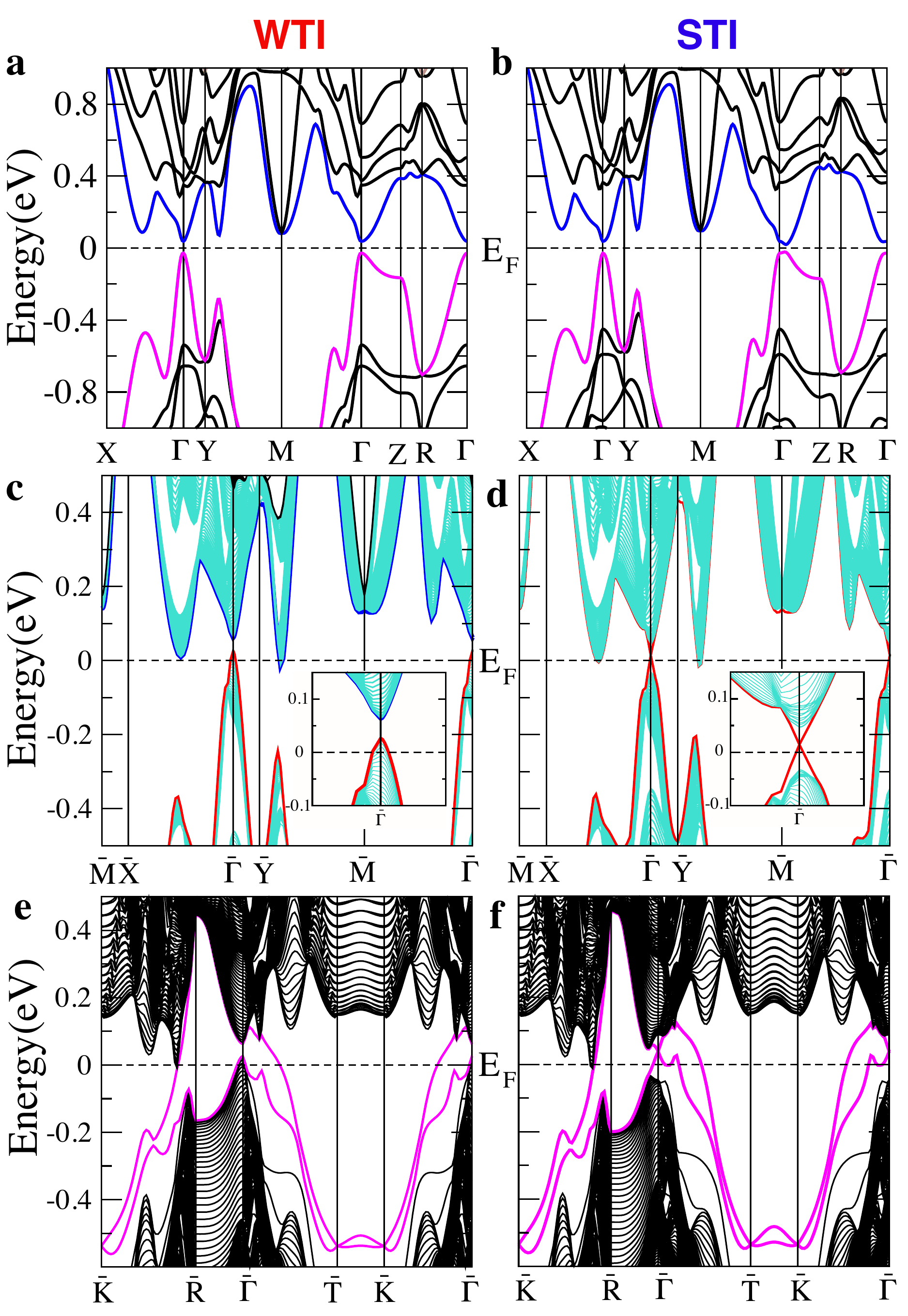}
 \caption{
    The calculated band structures and surface
    states for the 3D ZrTe$_5$. (a) and (b) are for the bulk, (c) and
    (d) are for the top surface, and (e) and (f) are for the side
    surfaces as indicated in Fig. 1. The weak TI (WTI) solutions (left
    panels) are obtained by using optimized lattice parameters, while
    the strong TI (STI) solutions (right panels) are from the
    experimental ones. (see main text for details).
    }
\end{figure}

\subsection{3D weak and strong TI} 
Given the QSH state in single layer ZrTe$_5$, it is important to check
if the stacked 3D compound is a weak or a strong
TI~\cite{Z2}. Interestingly, if we use the experimental lattice
parameters~\cite{structure}, we get a 3D strong TI with topological
invariant (1; 001); however, if the optimized lattice parameters,
which are only about 2\% larger than experimental ones, are used, we
get a 3D weak TI state\cite{weakstrong} with topological invariant (0; 001). The
difference between two solutions are due to the tiny overlap of the
states at the $\Gamma$ point of 3D BZ, which inverse the parity (see
Fig. 6). If we use the mBJ potential~\cite{mBJ} instead of GGA for the
above two calculations, we get both strong TI solutions. We have to
conclude that the required accuracy to distinguish these two states
for reality is beyond the ability of our present
calculations. Nevertheless, our results suggest that the 3D compound
is located very close to the boundary between the weak and strong TI,
which provides us an excellent opportunity to study the topological
phase transition experimentally. In Fig. 6, we show the calculated
band structures and the surface states of both strong and weak TI
solutions. Odd number of Dirac cones, one for the top- and three for
the side-surface, are obtained in the strong TI phase. For the weak TI
case, however, even number of Dirac cones, none for the top- and two
for the side surface, are obtained. It will be very interesting to
observe such topological phase transition in experiments just simply
pressing or stretching the sample along the layer stacking direction. It will be
also heuristic to see whether the long standing mysterious giant
resistivity anomaly~\cite{anomaly,anomaly-by-H} observed in both
ZrTe$_5$ and HfTe$_5$ has anything to do with the possible weak to
strong TI phase transition.

\section{Discussion}
Finally, we have performed all the calculations for HfTe$_5$ and draw
the same conclusions as ZrTe$_5$. Their 2D and 3D band structures and
size of band gap are all very similar. It
is experimentally known that the transition metal site Zr or Hf can be
substituted by many other elements, such as rare earth, which even
shows precursor of possible magnetic ordering~\cite{M-doping}. The
breaking of time reversal symmetry (TRS) will allow us to study the
possible quantum anomalous Hall effect~\cite{AHE,AHE-exp}. 
Moreover, due to the strong 2D features of these materials,
the edge states will  appear not only at the edge of the single-layer samples but
also at the terrace edges on the top surface of the 3D materials.  When such terrace edges
are covered by a s-wave superconducting layer, 1D topological superconductors
with Majorana bound states at the end of the terrace edges will be induced by 
the superconducting proximity effect\cite{Majorana2,Majorana3,Majorana1}.

\section{Acknowledgements}
We acknowledge the supports from the NSF of china and the 973 program of 
China (No. 2011CBA00108 and 2013CBP21700).

\section{ Appendix}

In this appendix, we prove that for single layer ZrTe$_5$ the energy bands at all the
zone boundary TRIM points are at least doubly degenerate with opposite parity. 
For simplicity, we neglect the spin degree of freedom and the conclusion can be 
easily generalized to the spin-1/2 case. 
Since the energy bands of ZrTe$_5$ near the Fermi level are formed by
Te 5$p$-orbitals, the Wannier functions centered at each atomic sites
with particular atomic feature (like $p_{x}$, $p_{y}$ and $p_{z}$)
carry a natural representation for the space group. First we define
the following Wannier basis in the momentum space,
\[
\phi_{\alpha
  k}^{\mu}(r)=\frac{1}{\sqrt{N}}\sum_{i}\phi_{\alpha}(r-\tau_{\mu}-R_{i})
  e^{ik(R_{i}+\tau_{\mu})}
  \], 
  where $i$ is the unit cell index, $\mu$ denotes the different atomic
  positions in a unit cell and $\alpha$ labels different orbitals on a
  same atomic position. When a general space group operator
  $\{P_{R}\mid t_{R}\}$, with $P_{R}$ and $t_{R}$ representing the
  rotation and translation part of the operator respectively, acts on
  the Wannier basis, we have
\[
\left\{ P_{R}\mid t_{R}\right\} \phi_{\alpha k}^{\mu}(r)
=t_{R}\left(P_{R}\phi_{\alpha k}^{\mu}(r)\right)
\]
\[
=\frac{1}{\sqrt{N}}\sum_{i'}\phi_{\mathcal{R}\alpha}(r-t_{\mathcal{R}}
-\mathcal{R}\tau_{\mu}-R_{i'})e^{i(\mathcal{R}k)(R_{i'}+\mathcal{R}\tau_{\mu})}
\]
\[
=\frac{1}{\sqrt{N}}\sum_{i'\alpha'\mu'}\phi_{\alpha'}(r-R_{0}^{\mu}
-\tau_{\mu'}-R_{i'})e^{i(\mathcal{R}k)(R_{i'}+R_{0}^{\mu}+\tau_{\mu'}-t_{R})}
\]
\[
=\phi_{\mathcal{R}\alpha,\mathcal{R}k}^{\mu'}(r)e^{-i(\mathcal{R}k)t_{R}}
=\sum_{\alpha'\mu'}Z_{\mu\mu'}^{\mathcal{R}}O_{\alpha\alpha'}^{\mathcal{R}}\phi_{\alpha',\mathcal{R}k}^{\mu'}(r)e^{-i(\mathcal{R}k)t_{R}}
\]
where we have take
$t_{R}+\mathcal{R}\tau_{\mu}=R_{0}^{\mu}+\tau_{\mu'}$ and
$\alpha'=\mathcal{R}\alpha$, $Z^{\mathcal{R}}$ and
$ $$O^{\mathcal{R}}$are matrices describing the transformation in
atomic positions and orbitals respectively.  We would emphasize that
in the above equation $\phi_{\alpha\mathcal{R}k}^{\mu}(r)$ does not
necessary equal to $\phi_{\alpha k}^{\mu}(r)$, even when
$\mathcal{R}k=k+G_{n}$ with $G_{n}$ representing a vector in reciprocal
lattice. They can differ by a phase factor due to the specific choice
of the gauge fixing condition here, which contains a phase factor
$e^{ik\tau_{\mu}}$ for orbitals located at different atomic position
$\tau_{\mu}$. Using the Wannier representation introduced above and
choosing the Zr site as the origin of the coordinate frame, we pick
two operators, inversion and mirror $m_{yz}$, which can be written as the
following
\[
\{I|\frac{1}{2},\frac{1}{2}\}\; and\;\{m_{yz}|0,0\}
\]

First we act $ $$\{I|\frac{1}{2},\frac{1}{2}\}\cdot\{m_{yz}|0,0\}$ to
the Wannier basis, which lead to
\[\{I|\frac{1}{2},\frac{1}{2}\}\centerdot\{m_{yz}|0,0\}\phi_{\alpha
  k}^{\mu}(r)=
\]
\[\left[Z^{I}Z^{yz}\right]_{\mu\mu'}\left[O^{I}O^{yz}\right]_{\alpha,\alpha'}\phi_{\alpha'(Im_{yz}k)}^{\mu'}(r)e^{-i(Im_{yz}k)t_{R}}
\]

where we have $t_{R}=\{\frac{1}{2},\frac{1}{2}\}$. And similarly when
we act $\{m_{yz}|0,0\}\cdot\{I|\frac{1}{2},\frac{1}{2}\}$, we have

\[
\{m_{yz}|0,0\}\centerdot\{I|\frac{1}{2},\frac{1}{2}\}\phi_{\alpha
  k}^{\mu}(r)
\]
\[=\left[Z^{yz}Z^{I}\right]_{\mu\mu'}\left[O^{yz}O^{I}\right]_{\alpha,\alpha'}\phi_{\alpha'(m_{yz}Ik)}^{\mu'}(r)e^{-i(Ik)t_{R}}
\]

Since we can easily prove that $Z^{yz}Z^{I}=Z^{I}Z^{yz}$ and
$O^{yz}O^{I}=O^{I}O^{yz}$, the only difference between the above two
is the phase factor, which differs by $-1$ for $(\pi,0)$ and
$(\pi,\pi)$$ $. Therefore for these two points we have
$\{I|\frac{1}{2},\frac{1}{2}\}\cdot\{m_{yz}|0,0\}=-\{m_{yz}|0,0\}\cdot\{I|\frac{1}{2},\frac{1}{2}\}$.
Because both the above symmetry operators commute with the
Hamiltonian, we can easily prove that if $\phi_{k}$ is eigen state of
Hamiltonian with given parity, the state $\{m_{yz}|0,0\}\phi_{k}$ is
another eigen state of Hamiltonian but with opposite parity. Therefore
we have proved that at $(\pi,0)$ and $(\pi,\pi)$, all the states come
in pairs with opposite parity. Replacing $m_{yz}$ by another mirror
$m_{xz}$, we can easily prove that another zone boundary TRIM point
$(0,\pi)$ has the same property. In conclusion, at all these zone boundary
TRIM points, the eigen states always come in degenerate pairs with
opposite parity.

\section{Author contributions}
H.M.W. did the material search and the first principle calculations. X.D. and Z.F.
did the symmetry and topology analysis.  All the authors contributed to the analysis
of the computational data and writing the manuscript.

\end{document}